\documentclass[aps, prl, twocolumn,showkeys, showpacs,amsmath,amssymb,superscriptaddress]{revtex4-1}
\usepackage{graphicx}
\usepackage{dcolumn}
\usepackage{bm}
\usepackage{epsfig}
\usepackage{subfigure}
\usepackage{color}
\newcommand{\be}{\begin{equation}}
\newcommand{\ee}{\end{equation}}
\newcommand{\bea}{\begin{eqnarray}}
\newcommand{\eea}{\end{eqnarray}}

\begin{document}
\title{A Gauge Condition for Studying the Origin of Intrinsic Magnetospheres}

\author{M. Mendoza} \email{mmendozaj@unal.edu.co} \affiliation{
  Departamento de F\'{\i}sica, Universidad Nacional de Colombia, Bogot\'a, D.C., (Colombia)}
\affiliation{ ETH
  Z\"urich, Computational Physics for Engineering Materials, Institute
  for Building Materials, Schafmattstrasse 6, HIF, CH-8093 Z\"urich
  (Switzerland)}

\author{J. Morales} \email{jmoralesa@unal.edu.co}
\affiliation{Associate researcher of Centro Internacional de
  F\'{\i}sica, Santaf\'e de Bogot\'a, (Colombia)}

\date{\today}
\begin{abstract}
  We propose an analytical model based on the solution of the magnetohydrodynamics (MHD) equations for studying the origin of intrinsic magnetospheres. For this purpose, we reveal a new gauge condition for the electromagnetic vector potential, which eases the solution of such complex system of non-linear equations. Using this model, we analyse the deformation of the terrestrial magnetic field due to the presence of the solar wind. By comparing with experimental observations, we have found that the geometrical configuration of the magnetosphere, before, and after the magnetic field of the Earth started to deviate the solar wind, has not changed notably, and that the solar wind should have had  a finite conductivity. This model can also be used to perform linear stability analysis of fluid and magnetic instabilities.
\end{abstract}


\maketitle

\begin{figure}
\includegraphics[width=0.8\columnwidth]{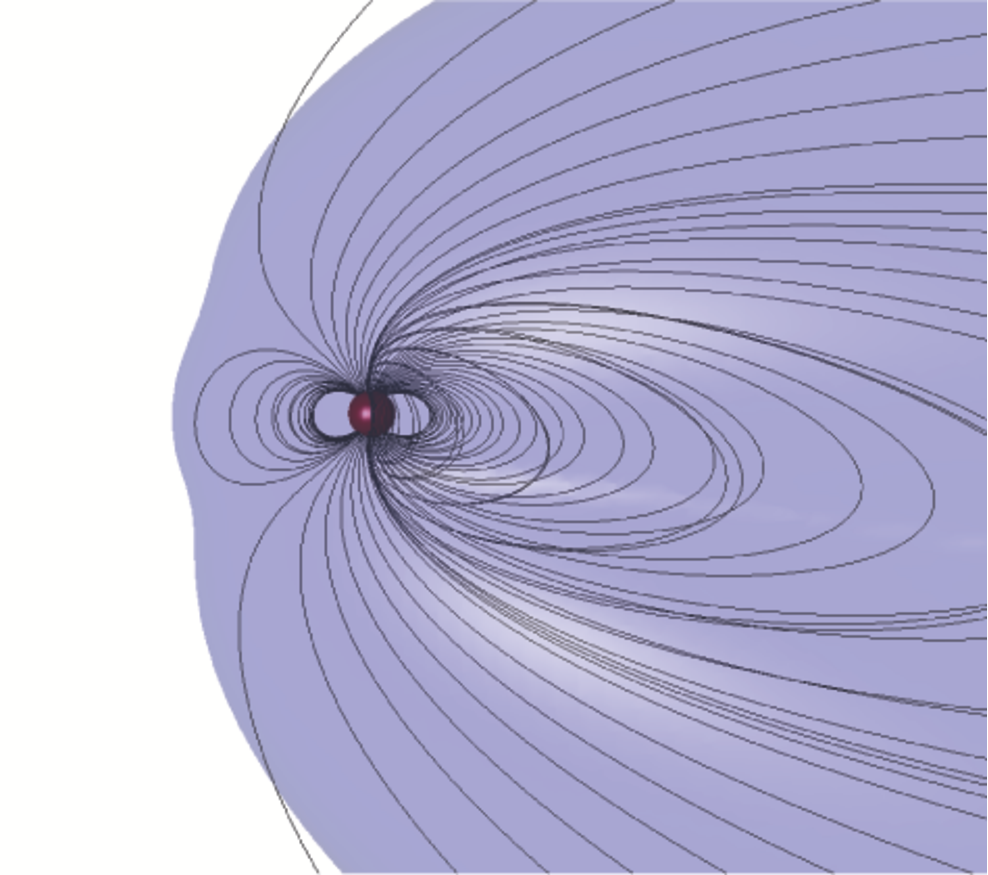}
\caption{Terrestrial magnetic field in the early stage deformed by the presence of the solar wind; the solar wind comes from left to right at $440$km/s. The red sphere represents the Earth and the blue iso-surface is located at $\xi \equiv |\vec{B}|/M = 2.56$ nT m$^2$/A.}\label{fig1a}
\end{figure}

The magnetosphere is a region near to an astronomical object where the surrounded plasma interacts with its magnetic field. In the case of the Earth, its magnetic field is deformed by the solar wind (see Fig.~\ref{fig1a}), which is a ``ionized gas'' with a very large electrical conductivity, chiefly made up of protons and electrons moving at high velocities \cite{magnetosphere}. The study of the terrestrial magnetosphere is very important, since it protects the Earth surface, satellites, telecommunication systems, and, in general, electric power grids, from the hot and highly conductive solar wind \cite{intro1,intro2}. 

Magnetic fields in the presence of plasmas are governed by the magnetohydrodynamics (MHD) equations. Finding steady-state solutions of these equations is very useful as a starting point to study magnetic and plasma instabilities. However, it is well known that analytical solutions for the MHD equations are rare, and for most practical problems the known solutions are not applicable. Although numerical solutions are relatively easy to achieve, for solving many theoretical and experimental problems in magnetospheric physics, magnetic field models are in general needed and can offer more understanding about the underlying physics. There are some quantitative models for the external geomagnetic field \cite{model1,model2,model3,model4,model5,model6,model7,model8}, which are widely used for various purposes for the magnetospheric community. At least two kind of models can be distinguished. The first class of models is based on experimental observations \cite{model1emp,model2emp,model3emp,model4emp,model5emp,model6emp}, and are constructed by minimising the discrepancy between model outputs and satellite observations. These models describe accurately the magnetosphere. A second class of models \cite{model4,model7,model3,model2,model9,modelul} uses a potential field approach in order to model the magnetic effect of the Chapman-Ferraro currents on the magnetopause. This second class of models can be used to assess magnetospheric configurations which differ significantly from today's Earth's magnetosphere. For examples, the potential field model approach was successfully used to model the magnetosphere of other planets \cite{uranus,neptuno}. 

The potential field approach is constructed on the basis of the shape of the magnetopause and the parameters of the internal field. Therefore, it considers the Chapman-Ferraro currents as input configurations instead of coming as theoretical predictions. This model is very good for studying the actual configuration of the magnetospheres based on observational data, but cannot study their origin since they assume a configuration that already exists. In fact, a theoretical model capable to reproduce the interaction between the solar wind and an intrinsic magnetosphere from first principles has never been developed before, to the best of our knowledge. In this Letter, we propose an analytical steady solution of the MHD equations for the case of an intrinsic magnetosphere, in the early stage when the magnetic field is not strong enough to deviate appreciably the solar wind. For this purpose, we introduce a new electromagnetic gauge condition that eases the solution of the MHD equations. By applying this model to the Earth and comparing with experimental observations, we find that the shape of the magnetosphere, before, and after the magnetic field of the Earth started to deviate the solar wind, has not changed appreciably. Furthermore, our solution shows that the solar wind should have had a finite conductivity. Other features, as the Chapman-Ferrato currents, are also observed.



We start our derivation by considering a highly conductive plasma in the presence of a magnetic field, e.g. the magnetic field generated by the Earth. For this case, we know that the magnetic field should satisfy the diffusion equation \cite{cowling,jackson},
\begin{eqnarray}\label{diffusion}
  \frac{\partial \vec{B}}{\partial t}=\eta_m \nabla^2 \vec{B} +
  \nabla \times (\vec{v}\times\vec{B}) \quad ,
\end{eqnarray}
where $\eta_m=1/\mu \sigma$ is the magnetic viscosity, $\sigma$
the conductivity, and $\mu$ the magnetic permeability of the plasma. Most of the geomagnetic models assume infinite conductivity of the solar wind, however, we will assume here that the conductivity is finite, and later show that if the plasma would have infinite conductivity, the terrestrial magnetosphere could not be formed.

The evolution of the plasma is given by solving the momentum conservation equation\cite{cowling,jackson},
\begin{eqnarray}\label{navier-stokes}
  \rho \left (\frac{\partial \vec{v}}{\partial t} + \vec{v} \cdot \nabla
    \vec{v}\right ) =-\nabla P+ \vec{j}\times
  \vec{B}+\eta \nabla^2 \vec{v} \quad ,
\end{eqnarray}
and the continuity equation,
\begin{eqnarray}\label{continuity}
  \nabla \cdot (\rho \vec{v})=-\frac{\partial \rho}{\partial t} \quad .
\end{eqnarray}
Here, $\rho$ is the density and $\vec{v}$ the velocity,
$P$ the hydrostatic pressure, $\eta$ the shear viscosity, and $\vec{j}$ the
electric density current. The solar wind is nearly inviscid, and
therefore in this study we will consider $\eta = 0$. Furthermore, we assume that the magnetic field generated by the astronomical object is not strong enough to deviate appreciably the solar wind, i.e. the plasma parameter
$\beta \equiv 2\mu P/B^2 \gg 1$, and therefore, the forcing term
$\vec{j}\times \vec{B} \propto B^2$ can be neglected. Thus,
Eqs.~\eqref{continuity} and \eqref{navier-stokes} are a complete set
of differential equations describing the dynamics of the solar wind,
and are decoupled from Eq.~\eqref{diffusion} for the magnetic
field. Note that the opposite is not true, the magnetic diffusion
equation depends on the velocity field.

We study the case where density and velocity fluctuations in the
solar wind due to the magnetic field of the source are negligible, so
that, as a first approximation, we consider an incompresible ($\nabla
\cdot \vec{v} = 0$) and irrotational ($\nabla \times \vec{v} = 0$)
fluid in the steady state, $\partial /\partial t = 0$. By doing this,
we are excluding the dynamics related to the formation of vortices,
that is generally formed in the proximities of the magnetic
source, and also magnetic reconnection processes \cite{priest2007,reconnection1,mhd_miller}. According to these considerations, the velocity of the fluid can be written as the gradient of a potential function $\phi_v$, such that
\begin{equation}
  \vec{v}=-\nabla \phi_v \rightarrow \nabla^2 \phi_v=0 \quad .
\end{equation}

On the other hand, by using the vector potential $\vec{A}$, such that $\vec{B}=\nabla \times \vec{A}$, and using the vectorial relation,
\begin{eqnarray}
  \nabla \times (\nabla \times \vec{A})=\nabla(\nabla \cdot
  \vec{A})-\nabla^2\vec{A}. \label{2,6}
\end{eqnarray}
we can write Eq.~\eqref{diffusion} as
\begin{eqnarray}\label{diff-A}
  \eta_m \left [\nabla (\nabla \cdot \vec{A}) - \nabla^2 \vec{A}
  \right ] = \vec{v}\times (\nabla
  \times \vec{A}) \quad . 
\end{eqnarray}

By keeping in mind that the fluid is irrotational and the velocity fluctuations are negligible, the identity relation
\begin{eqnarray}
  \begin{aligned}
    \nabla (\vec{v}\cdot \vec{A})=(\vec{v}\cdot
    \nabla)\vec{A}&+(\vec{A}\cdot \nabla)\vec{v}\\ &+\vec{v}\times
    (\nabla \times \vec{A})+\vec{A}\times (\nabla \times \vec{v})
    \quad ,
  \end{aligned}
\end{eqnarray}
can be written as
\begin{eqnarray}
  \vec{v}\times (\nabla \times \vec{A})=\nabla(\vec{v}\cdot
  \vec{A})-(\vec{v}\cdot \nabla)\vec{A} \quad . 
\end{eqnarray}
Replacing this equation into Eq.~\eqref{diff-A}, we get
\begin{eqnarray}\label{eq-solv}
  \eta_m \nabla^2 \vec{A}-(\vec{v}\cdot \nabla)\vec{A}=\nabla
  (\eta_m \nabla \cdot \vec{A}-\vec{v}\cdot \vec{A}) \quad .
\end{eqnarray}

This equation must be solved in order to know the deformation of the magnetic field
due to the interaction with the solar wind. However, solving this
equation is not simple, and here we introduce an unconventional way to
do it. We will first set the gauge condition $\eta_m \nabla \cdot
\vec{A}-\vec{v}\cdot \vec{A} = 0$, and find a solution to the
equation,
\begin{eqnarray}\label{eq-solv1}
  \eta_m \nabla^2 \vec{A}-(\vec{v}\cdot \nabla)\vec{A}=0 \quad ,
\end{eqnarray}
for the vector potential $\vec{A}$, and afterwards, we use the constraint to find the real solution, including the integration parameters. Thus, replacing $\vec{v} = -\nabla \phi_v$ into Eq.~\eqref{eq-solv1}, we obtain
\begin{eqnarray}\label{eq-solv3}
  \eta_m \nabla^2 \vec{A} + \nabla \phi_v \cdot \nabla \vec{A} = 0
  \quad .
\end{eqnarray}
Let us now propose a solution with the following form:
\begin{eqnarray}\label{assum1}
\vec{A} = \phi_m \vec{A}_0 \quad , 
\end{eqnarray}
such that, if we replace it into Eq.~\eqref{eq-solv3}, we obtain the
following relation:
\begin{eqnarray}
  \begin{aligned}
    \eta_m \phi_m \nabla^2 \vec{A}_0 &+ (\eta_m \nabla^2 \phi_m +
    \nabla\phi_v \cdot \nabla\phi_m) \vec{A}_0 \\ &+ (2\eta_m \nabla
    \phi_m + \phi_m \nabla \phi_v) \cdot \nabla \vec{A}_0 = 0 \quad .
  \end{aligned}
\end{eqnarray}
Note that by choosing,
\begin{eqnarray}\label{sol1}
\nabla \phi_m= -\frac{\phi_m}{2\eta_m}\nabla \phi_v \quad ,
\end{eqnarray}
we get the modified Helmholtz's equation,
\begin{eqnarray}\label{helm}
  \nabla^2 \vec{A}_0 - k^2 \vec{A}_0 =0 \quad .
\end{eqnarray}
where
\begin{eqnarray}\label{k-definition}
  k=\frac{1}{2\eta_m} \sqrt{\nabla\phi_v \cdot \nabla
    \phi_v}=\frac{\mu \sigma}{2} |\nabla\phi_v |
  \quad .
\end{eqnarray}

The solution of Eq.~\eqref{helm}, in spherical coordinates $(r,
\theta, \phi)$, is given in terms of the spherical harmonics,
$Y_{lm}(\theta, \phi)$, and the modified spherical Bessel polynomials,
$i_l(kr)$ and $n_l(kr)$ \cite{arfken}. Therefore, we can write the solution as
\begin{eqnarray}\label{sol3}
  \vec{A}_0=\sum_{l=0}^{\infty}\sum_{m=-l}^l \left[\vec{D}_{lm}n_l (kr)
  \right]Y_{lm}(\theta, \phi) \quad ,
\end{eqnarray}
where we have suppressed the contribution of $i_l(kr)$ since we require a converging solution for $r \rightarrow \infty$. On the other hand, the solution of Eq.~\eqref{sol1} can be obtained by
direct integration,
\begin{eqnarray}
  \phi_m = C_1 e^{-\phi_v/2 \eta_m} \quad . \label{sol2}
\end{eqnarray}

Replacing both solutions, Eqs.~\eqref{sol3} and \eqref{sol2}, into
Eq.~\eqref{assum1}, we can write the general solution for the
potential vector as
\begin{eqnarray}\label{multi_sol}
  \vec{A} =e^{-\phi_v/2\eta_m} \sum_{l=0}^{\infty}\sum_{m=-l}^l \left[\vec{D}_{lm}n_l (kr)
  \right]Y_{lm}(\theta, \phi). \label{sol_tot}
\end{eqnarray}
Note that we have redefined the integration constant $C_1 \vec{D}_{lm}
\rightarrow \vec{D}_{lm}$ for simplicity.

From now on, we assume that the source of the magnetic field is a
magnetic dipole. Therefore, the only remaining terms in
Eq.~\eqref{sol_tot} are
\begin{eqnarray}
  \vec{A} = e^{-\phi_v/2\eta_m - kr} \frac{M' (1 + kr)}{k^{3/2} r^3}
  (-y, x, 0) \quad ,
\end{eqnarray}
where $M'$ is a constant that will be related with the magnetic dipole moment. Note that one can also assume other kind of multi-polar expansion for more general cases. In order to have a valid solution of Eq.~\eqref{eq-solv}, we have imposed the gauge condition,
\begin{eqnarray}\label{gauge}
  \eta_m \nabla \cdot \vec{A} + \nabla \phi_v \cdot \vec{A} =0 \quad ,
\end{eqnarray}
which must be satisfied. For this purpose, we can include additional terms in the expansion of the solution, Eq.~\eqref{sol_tot}, obtaining
\begin{widetext}
  \begin{eqnarray}\label{solution}
    \vec{A} = M' e^{-\phi_v/2\eta_m - kr}
    \biggl (-\frac{y(1 + kr)}{k^{3/2} r^3}  + \frac{1}{\sqrt{k} r}\frac{1}{|\nabla \phi_v|}
    \frac{\partial \phi_v}{\partial y}, \frac{x (1 + kr)}{k^{3/2} r^3}
    - \frac{1}{\sqrt{k} r} \frac{1}{|\nabla \phi_v|}\frac{\partial \phi_v}{\partial x}, 0 \biggr ) \quad .
  \end{eqnarray}
\end{widetext}

The constant $M'$ is related with the magnetic dipole moment of the source, $M$. Since for low plasma velocity we expect to recover the solution of the vector potential for a magnetic dipole, we can conclude that $M' = k^{3/2} \mu M/4\pi$. Replacing this into Eq.~\eqref{solution}, and taking into account the definition of $k$, Eq.~\eqref{k-definition}, we obtain the final solution,
\begin{widetext}
  \begin{eqnarray}\label{solution_final}
    \vec{A} = \frac{\mu M}{4\pi} e^{-\phi_v/2\eta_m - kr}
    \biggl (-\frac{y(1 + kr)}{r^3}  + \frac{\mu \sigma}{2r}
    \frac{\partial \phi_v}{\partial y}, \frac{x (1 + kr)}{r^3}
    - \frac{\mu \sigma}{2r}\frac{\partial \phi_v}{\partial x}, 0 \biggr ) \quad .
  \end{eqnarray}
\end{widetext}
Thus, for $|\nabla \phi_v| \rightarrow 0$ or $\sigma \rightarrow 0$ (both cases imply $k \rightarrow 0$), we get the vector potential for a magnetic dipole. On the other hand, if the plasma posses very high conductivity, we see that the solution decays exponentially to zero, implying that at infinite conductivity, the formation of the magnetosphere could not be possible. From Eq.~\eqref{solution_final}, we can determine the magnetic field using the expression $\vec{B} =\nabla\times \vec{A}$.
Note that our model is able to describe more general cases, e.g. multipolar terms, by including more terms in the expansion in Eq.~\eqref{multi_sol}. However, we must always satisfy the gauge condition given by Eq.~\eqref{gauge}, or less strict gauge condition, $\nabla (\eta_m \nabla \cdot \vec{A}-\vec{v}\cdot \vec{A}) = 0$. Furthermore, during our calculations we have always considered constant velocity, $\vec{v} = -\nabla \phi_v$, but in principle, we could also study small velocity fluctuations. 

\begin{figure}
\includegraphics[width=\columnwidth]{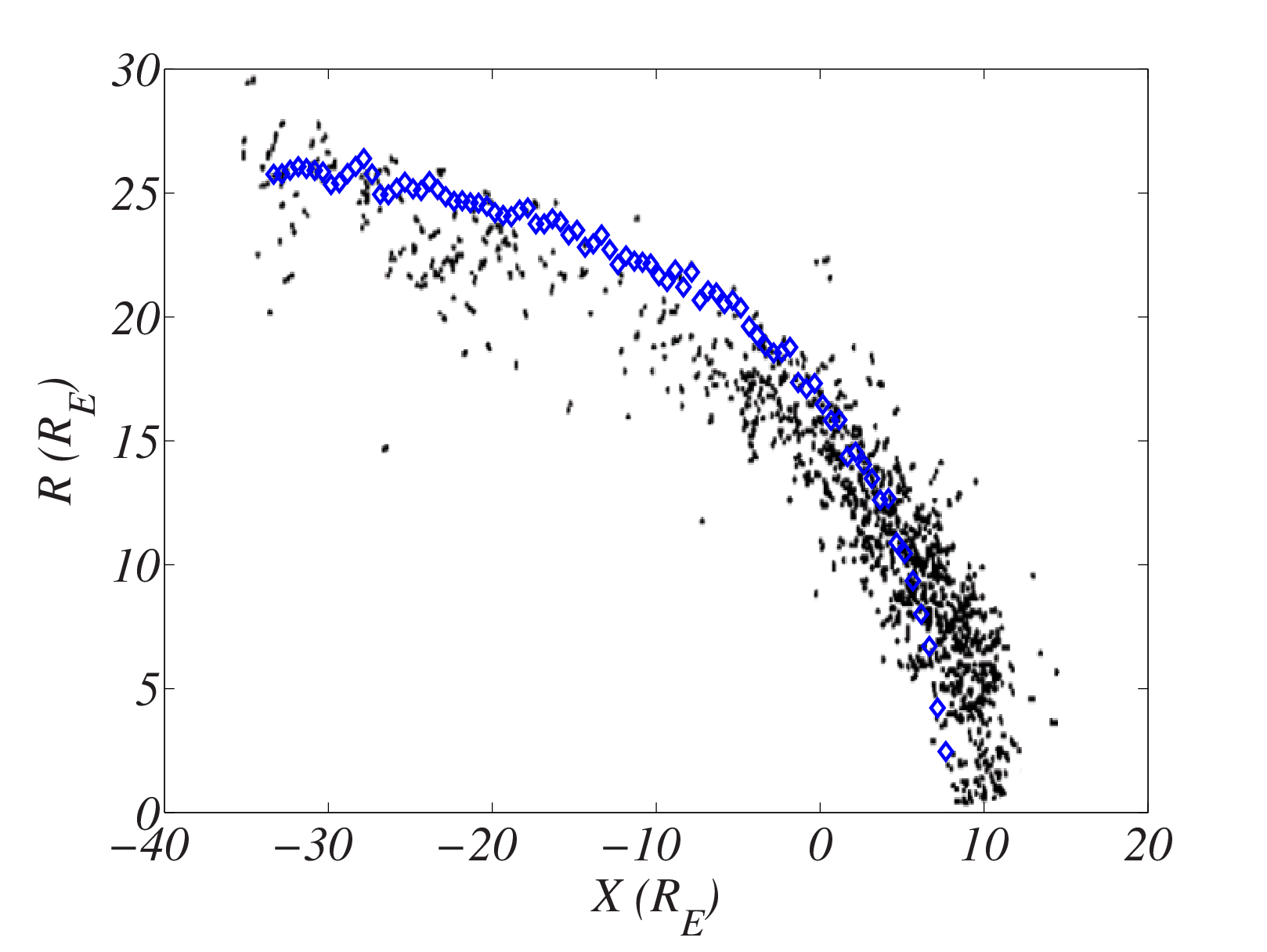}
\caption{Location and shape of the magnetopause. The black points are experimental data \cite{magnetopause_ref}, and the blue diamonds represent our theoretical results for the region $\xi = 2.56$ nT m$^2$/A. $R_E$ is the Earth radius and $R = \sqrt{Y^2 + Z^2}$, being the magnetopause located at $(X,Y,Z).$}. \label{fig2}
\end{figure}
\begin{figure}
\includegraphics[width=\columnwidth]{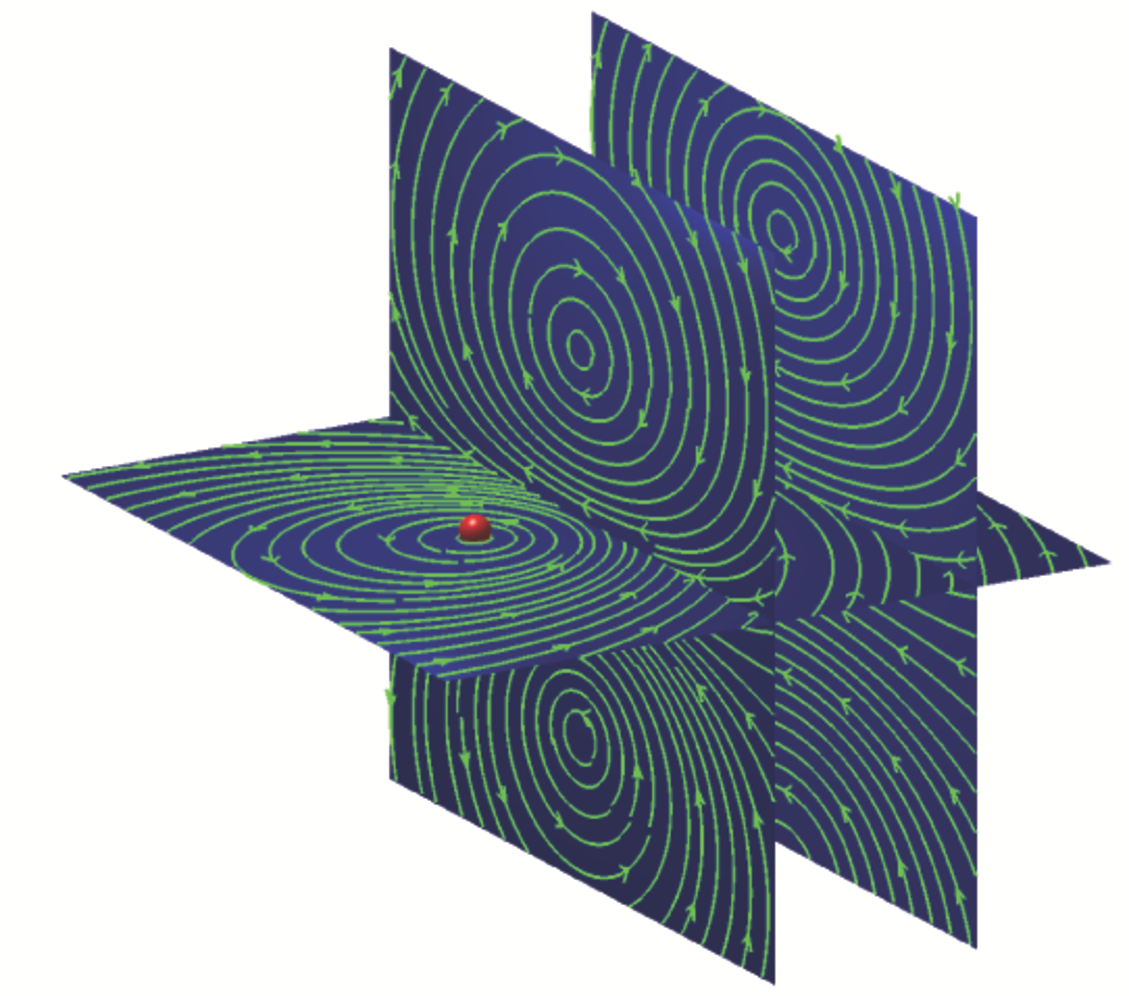}
\caption{Induced electrical current density in the solar wind due to the presence of the terrestrial magnetic field. The solar wind comes from left to right at $440$km/s. The green streamlines show the current density lines and the red sphere denotes the Earth. The electrical currents in the vertical planes are also known as Chapman-Ferraro currents.}\label{fig1b}
\end{figure}

From the analytical expression, Eq.~\eqref{solution_final}, is difficult to see the configuration of the magnetic field. Therefore, we replace the characteristic values from the terrestrial magnetosphere and study this specific case. We have taken for the magnetic dipole moment $M = 7.8\times 10^{22}$ A/m$^2$ \cite{dipole_ref}, solar wind conductivity $\sigma \simeq 10^{-6}$ S/m \cite{sigma_ref}, and a solar wind speed of $v = 440$ m/s \cite{speed_ref}. Since the magnetic dipole is tilted around $\lambda \simeq 11$ degrees \cite{dipole_ref}, we set up a velocity vector of the form $\vec{v} = v (\cos(\lambda), 0, -\sin(\lambda))$, and its corresponding velocity potential $\phi_v$, keeping the magnetic dipole in the direction $z$. In Fig.~\ref{fig1a}, we can observe the magnetic lines configuration, which qualitatively looks very similar to the real magnetospheric configuration. The blue iso-surface is located at $\xi \equiv |\vec{B}|/M = 2.56$ nT m$^2$/A, which corresponds approximately to the magnitude of the magnetic field at the observed magnetopause \cite{magnetopause_ref}. As a matter of comparison, we have also measured the shape of this region predicted by the theory, finding very good agreement (see Fig.~\ref{fig2}). That is very surprising since we do not assume a magnetopause nor a solar wind deviation due to the terrestrial magnetic field. This suggests that the shape of this region was already set even before the terrestrial magnetic field was able to deviate the solar wind and create the actual magnetopause. It also implies that, after the magnetopause formation and the emergence of the strong deviation of the solar wind due to the terrestrial magnetic field, this shape has changed very little. The Chapman-Ferraro currents are also reproduced at least qualitatively in Fig.~\ref{fig1b}. The current density can be calculated from Eq.~\eqref{solution_final} by applying the curl twice, $\vec{J} = \nabla\times (\nabla \times \vec{A}) / \mu$. Here we see that the complex configuration of these currents was present during the formation of the terrestrial magnetic field, which are the responsible for the appearance of the magnetopause.

In summary, we have found an analytical solution of the MHD equations to study the origin of intrinsic magnetospheres. In order to achieve such solution, we have introduced a new gauge condition such that the analytical solution of the equations is possible. By replacing actual values of the solar wind and the terrestrial magnetic field, we have found that our model is able to reproduce the shape of the actual magnetopause and also describe qualitatively the presence of the Chapman-Ferraro currents. This surprising result suggests that before the formation of the magnetopause and the emergence of the strong deviation of the solar wind due to the terrestrial magnetic field, this shape has remained unchanged. 

The theoretical model proposed here, can be applied to other kind of planets by choosing different multipolar expansions of the solution of the MHD equations. Temporal perturbations and MHD instabilities can also be studied using our steady solution and will be subject of future research. Extensions to consider plasma deviations and the formation of the magnetopause will also be considered in future works.

%

\bibliography{mhd}

\end{document}